\documentclass[superscriptaddress,showpacs,aps,twocolumn,prb,floatfix]{revtex4-1}
\usepackage{amssymb}
\usepackage{amsmath}
\usepackage[dvips]{graphicx}
\usepackage{bm}

\newcommand{\bea}{\begin{eqnarray}}
\newcommand{\eea}{\end{eqnarray}}

\def\beq{\begin{equation}}
\def\eeq{\end{equation}}

\begin{document}

\title{Nonequilibrium conductance through a benzene molecule in the Kondo regime}
\author{L. Tosi}
\affiliation{Centro At\'omico Bariloche and Instituto Balseiro, Comisi\'on Nacional de
Energ\'{\i}a At\'omica, 8400 Bariloche, Argentina}
\author{P. Roura-Bas}
\affiliation{Dpto de F\'{\i}sica, Centro At\'{o}mico Constituyentes, Comisi\'{o}n
Nacional de Energ\'{\i}a At\'{o}mica, Buenos Aires, Argentina}
\author{A. A. Aligia}
\affiliation{Centro At\'omico Bariloche and Instituto Balseiro, Comisi\'on Nacional de
Energ\'{\i}a At\'omica, 8400 Bariloche, Argentina}

\date{\today }

\begin{abstract}
Starting from exact eigenstates for a symmetric ring, we derive a low-energy
effective generalized Anderson Hamiltonian which contains two spin doublets
with opposite momenta and a singlet for the neutral molecule. For benzene,
the singlet (doublets) represent the ground state of the neutral (singly
charged) molecule. We calculate the non-equilibrium conductance through a
benzene molecule, doped with one electron or a hole (i.e. in the Kondo
regime), and connected to two conducting leads at different positions. We
solve the problem using the Keldysh formalism and the non-crossing
approximation (NCA). When the leads are connected in the \emph{para}
position (at 180 degrees), the model is equivalent to the ordinary impurity
Anderson model and its known properties are recovered. For other positions,
there is a partial destructive interference in the cotunneling processes
involving the two doublets and as a consequence, the Kondo temperature and
the height and width of the central peak (for bias voltage $V_b$ near zero) of
the differential conductance $G=dI/dV_b$ (where $I$ is the current) are
reduced. In addition, two peaks at finite $V_b$ appear. We study the
position of these peaks, the temperature dependence of $G$ and the spectral
densities. Our formalism can also be applied to carbon nanotube quantum
dots with intervalley mixing.
\end{abstract}

\pacs{73.23.-b,73.22.-f, 75.20.Hr}
\maketitle

\section{Introduction}
\label{intro}

Single molecule electronic devices are being extensively studied because
they offer perspectives for further miniaturization of electronic circuits
with important potential applications \cite{nitzan,venk,galp,molen,cuevas},
and also due to its intrinsic interest in basic research, for example as
realizations of the Kondo effect, which manifests experimentally by an
increased conductance at low temperatures \cite{park,lian}. In addition,
quantum phase transitions involving partially Kondo screened spin-1
molecular states were induced changing externally controlled parameters 
\cite{roch,parks,serge}. These experiments could be explained semiquantitatively
using extensions of the impurity Anderson model treated with either the
numerical renormalization group (NRG) \cite{parks,serge,epl} or with the
non-crossing approximation (NCA) \cite{serge,epl,st1,st2}. This approximation
allows calculations out of equilibrium and in particular at finite bias
voltage.

The conductance through benzene-1,4-dithiol has been measured using a
mechanically controllable break junction \cite{reed}. 
More recently, other molecules containing benzene rings or related phenyl groups were 
studied \cite{dani,dado}.
The states nearer to
the Fermi level in benzene are built from the 2p states of the C atoms which
lie perpendicular to the plane of the molecule (the so called $\pi $
states). Transport through aromatic molecules and in particular through
benzene in different geometries has been calculated by several 
groups \cite{hett,carda,ke,bege,mole}. Hettler \textit{et al.} \cite{hett}  
started from
an exact calculation of an extended Hubbard model for the $\pi $ states, and
included the coupling to the leads perturbatively (for couplings smaller
than the temperature), to calculate the current through neutral benzene
under an applied bias voltage $V_b$ for zero gate voltage $V_{g}$, including
radiative relaxation. Cardamone \textit{et al.} \cite{carda}, described the
molecule by a Hubbard model supplemented by repulsions at larger distance
(the so called Pariser-Parr-Pople (PPP) Hamiltonian \cite{ppp}) using the
self-consistent Hartree-Fock approximation. The conductance of the effective
one-body problem for small $V_{g}$ has been calculated using the 
Landauer-B\"{u}ttiker formalism \cite{but}, The authors propose to exploit the
destructive interference and to control the conductance of the device
introducing symmetry breaking perturbations. Similar results were obtained
using \textit{ab initio} calculations \cite{ke}, which have also the drawback
of neglecting the effect of correlations.

This effect together with that interference were included in more recent
works which started from the exact solution of the PPP model \cite{bege,mole},
as well as in previous works which discussed the conductance through
molecules or rings of quantum dots (QDs) threaded by a magnetic flux 
\cite{jagla,hall,ihm,frie,soc,rinc}. The conductance $G$ depends on which sites of the
molecule are coupled to the leads and on interference phenomena related to
the symmetry of the system \cite{bege,mole,rinc}. For certain conditions $G$
can be totally suppressed and restored again by symmetry-breaking
perturbations \cite{mole}. While the main conclusions are safe in general,
the coupling to the leads was included in some perturbative approach, 
either by a Liouville equation method \cite{bege}, or
using the Jagla-Balseiro formula (JBF) for the conductance \cite{jagla,ihm},
which assumes a non-degenerate ground state. These approaches miss the Kondo
effect which is non-perturbative in the coupling to the leads \cite{soc}.
This is particularly important when the gate voltage is such that the
benzene molecule becomes charged, because the conductance is larger \cite{mole}. 

The JBF has also been used to predict that the transmittance
integrated over a finite energy window \cite{jagla,frie,hall} and the
equilibrium conductance \cite{hall,rinc} through a ring of strongly
correlated one-dimensional systems display dips as a function of the applied
magnetic flux at \emph{fractional} values of the flux quantum, due to
destructive interference at crossings of levels with different charge and
spin quantum numbers \cite{rinc} (as a consequence of spin-charge
separation). Recently, we have confirmed the presence of dips in the current
under a finite applied bias voltage at low temperatures, using the
low-energy effective Hamiltonian, consisting of two doublets and a singlet,
for the case of perfect destructive interference \cite{desint}. 
At equilibrium (zero bias voltage), the conductance as a function of temperature
for this particular case has been calculated with NRG \cite{izum2}.
The interplay between
interference and interactions were also studied for spinless electrons in
multilevel systems \cite{meden}, and benzene attached to two leads \cite{bohr}.
Although at first sight they might seem artificial, spinless models describe
effective Hamiltonians for realistic systems under applied magnetic fields \cite{nils,pss}.

Interference phenomena were observed in systems with QDs. In a multilevel
QD, the crossing of levels and the ensuing destructive interference has been
induced applying magnetic field to a system with large, level-dependent $g$
factors \cite{nils}. Aharonov-Bohm oscillations were observed in systems
involving two quantum dots (QDs) \cite{hata}.

In this work, we construct the low-energy effective Hamiltonian for a ring
of $n$ sites with one orbital per site and symmetry that includes the point 
group $C_{nv}$ (or $C_{n}$), weakly coupled
to two conducting leads, retaining for $n$ even, the lowest singlet with $n$
particles and the two lowest doublets with $n+1$ particles (electrons or
holes depending on the sign of the applied gate voltage $V_{g}$). For $n$
odd, the charge of doublets and singlets are interchanged. Using a gauge
transformation, three of the four hopping matrix elements between the
doublets and the leads can be made real. The phase of the fourth one $\phi $
is in general different from zero and depends on the position of the leads
and the wave vectors of the states involved. While the coupling to the leads
should be small in such a way that the neglected states do not affect the
calculations \cite{lobos}, we treat this coupling in a self-consistent, non
perturbative way, using the the Keldysh formalism of the NCA for problems
out of equilibrium \cite{st2,nca,nca2}, appropriately extended for this
problem. We calculate the non-equilibrium conductance for case of benzene 
($n=6$) in a regime of $V_{g}$ for which the doublets are favored, leading to
Kondo effect.

For the case of one doublet, comparison of NCA with NRG results \cite{compa},
shows that the NCA describes accurately the Kondo physics. The leading
behavior of the differential conductance for small voltage and temperature 
\cite{roura} agrees with alternative Fermi-liquid approaches \cite{ogu,scali},
and the temperature dependence of the conductance practically coincides with
the NRG result over several decades of temperature \cite{roura}. A
shortcoming of the NCA is that, at very low temperatures, it introduces an artificial spike at the
Fermi energy in the spectral density when the ground state of the system
without coupling to the leads (zero hybridization) is non-degenerate,
although the thermodynamic properties continue to be well described \cite{nca}. 
This is not important in the present work, where the doublets are
below the singlet and there is no applied magnetic field. Another limitation
of the method is that it is restricted to temperatures above $\sim T_{K}/20$, 
where $T_{K}$ is the Kondo temperature. In our case, this limitation is
not important either because as we shall see, the conductance has already
saturated at temperatures above this limit. A virtue of the NCA that becomes
important in our case, is its ability to capture features at high energies,
such as peaks in the spectral density out of the Fermi level, which might be
lost in NRG calculations \cite{vau}. An example is the plateau at
intermediate temperatures observed in transport through C$_{60}$ molecules
for gate voltages for which triplet states are important \cite{roch,serge},
which was missed in early NRG studies, but captured by the NCA \cite{st1,st2}.
More recent NRG calculations using tricks to improve the resolution \cite{freyn}, 
have confirmed this plateau \cite{serge}.

For the particular case of perfect destructive interference $\phi =\pi $
(which does not correspond to benzene), the spectral densities \cite{fcm}
and the non-equilibrium current \cite{desint}, were calculated before for
symmetric coupling to the leads. In this case, the model is equivalent to an
SU(4) impurity Anderson model, used by several authors \cite{lim,ander,lipi,buss} 
to interpret transport experiments in C nanotubes 
\cite{jari,maka}, and more recently in Si fin-type field effect transistors \cite{tetta}.
If in addition a finite splitting $\delta $ between the
doublets is allowed, the symmetry is reduced to SU(2). The transition between
SU(4) and SU(2) has been studied by several authors using the NCA \cite{desint,fcm,lipi,tetta}.
In particular we found that the reduction of the
Kondo temperature with $\delta $ agrees very well with a simple formula
obtained from a variational calculation \cite{desint,fcm}. An important
difference between our model and those used for other systems is the
connectivity with the leads, which imposes that in our case, the equilibrium
conductance $G(V_b=0)$ is zero in the SU(4) limit $\delta =0$, and therefore,
it is not proportional to the total spectral density of states \cite{desint}.
For $\phi =\pi $ and any $\delta $, it is possible to calculate $G(0)$ in
terms of static occupations, using a generalized Friedel-Langreth sum 
rule \cite{desint}, due to the simplicity of the ``orbital" field $\delta $, which
acts in the same way as a magnetic field, and allows to relate the phase
shift for each channel and spin with the respective occupancies. In turn, these 
phase shifts enter a  
general Fermi-liquid expression for the conductance obtained
by Pustilnik and Glazman \cite{pus}. This is not possible for general $\phi$. 
Therefore a novel treatment is needed to calculate the conductance even at
equilibrium. 

Recently, it has been shown that in carbon nanotube QDs 
with disorder induced valley mixing, the SU(4) symmetry is broken and 
the tunnel couplings to metallic leads become complex and depend on the 
applied magnetic field \cite{grove}. The present formalism is also
suitable to study this case.

\begin{figure}[tbp]
\includegraphics[width=8.0cm ]{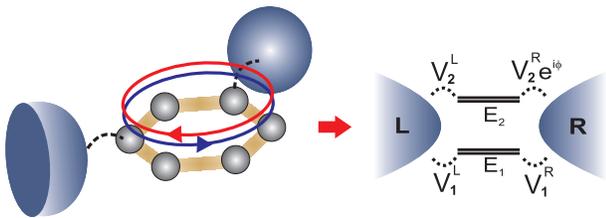}
\caption{(Color online) Scheme of the relevant matrix elements at low energies.}
\label{scheme}
\end{figure}

The paper is organized as follows. In Section \ref{model} we describe the construction
of the effective Hamiltonian. The approximations and the equation for the current are
presented in Section \ref{forma}. Section \ref{res} contains the numerical results
for transport through the benzene molecule and their interpretation. Section
\ref{sum} contains a summary and a short discussion. In appendix \ref{cc} we show that the NCA conserves 
the current, 
extending the existing demonstration for the case of one doublet \cite{nca}.
Appendix \ref{deta} contains some numerical details. 

\section{Model}
\label{model}

The effective model, which is represented at the right of Fig. \ref{scheme}, contains a
singlet with total wave vector $K_{0}$ (usually 0 or $\pi$) and two doublets with wave vectors 
$K_{1}$ and $K_{2}$, that represent the states of lowest energy of two
neighboring configurations of a ring of $n$ sites with symmetry $C_{nv}$. In
the case of benzene, they correspond to the singlet ground state, invariant
under rotations ( $K_{0}=0$) and two degenerate doublets with total wave
vectors $\pm K$, which are the ground state of the molecule for one added
electron or hole, depending on the sign of the applied gate voltage \cite{bege,mole,rinc}.
For one added hole $K=\pi /3$, while for one added electron 
$K=2\pi /3$. The effective Hamiltonian is 
\begin{eqnarray}
H &=&E_{s}|s\rangle \langle s|+\sum_{i\sigma }E_{i}|i\sigma \rangle \langle
i\sigma |+\sum_{\nu k\sigma }\epsilon _{\nu k}c_{\nu k\sigma }^{\dagger
}c_{\nu k\sigma }  \nonumber \\
&&+\sum_{i\nu k\sigma }(V_{i}^{\nu }|i\sigma \rangle \langle s|c_{\nu
k\sigma }+\mathrm{H.c}.),  \label{ham}
\end{eqnarray}%
where the singlet $|s\rangle $ and the two doublets 
$|i\sigma \rangle $ ($i=1,2$; $\sigma =\uparrow $ or $\downarrow $) denote the localized states, 
$c_{\nu k\sigma }^{\dagger }$ create conduction states in the left ($\nu =L$)
or right ($\nu =R$) lead, and $V_{i}^{\nu }$ describe the hopping elements
between the leads and both doublets, assumed independent of $k$. This
hopping element is calculated as \cite{rinc}

\begin{equation}
V_{i}^{\nu }=t_{\nu }\langle i\sigma |c_{j_{\nu }\sigma }^{\dagger
}|s\rangle ,  \label{hop}
\end{equation}
where $c_{j\sigma }^{\dagger }$ creates an electron (or hole) with spin $\sigma $ at the $\pi $ 
orbital at site $j$ (1 to $n$) of the molecule, $j_{\nu }$ denotes the site connected to the lead $\nu $, and $t_{\nu }$ is
the hopping between this site and the lead $\nu $.

Choosing adequately the phases of the gauge transformation $|i\sigma \rangle
\rightarrow e^{i\varphi _{i}}|i\sigma \rangle $, both $V_{i}^{L}$ can be
made real and positive, and by reflection symmetry 
$V_{1}^{L}=V_{2}^{L}$.
Using rotational symmetry, it is easy to see that \cite{rinc}

\begin{equation}
V_{i}^{R }=(t_{R}/t_{L})V_{i}^{L }\exp \left[ -i(j_{R}-j_{L})(K_{i}-K_{0})\right] ,  
\label{hopr}
\end{equation}
where $K_{i}$ is the wave vector of $|i\sigma \rangle $. 
The phase of $V_{1}^{R}$ can be absorbed by a gauge transformation in the $c_{Rk\sigma }$,
rendering it real and positive. The remaining matrix element is $V_{2}^{R}=V_{1}^{R}e^{-i\phi }$, 
where 
$\phi =(j_{R}-j_{L})(K_{2}-K_{1})$.
This result is general for a ring geometry.

For benzene $K_{2}-K_{1}\equiv \pm 2\pi /3$. Then, if the leads are connected
in the \emph{para} position ($j_{R}-j_{L}=3$) , $\phi \equiv 0$, while in
the \emph{ortho} ($j_{R}-j_{L}=1$) or \emph{meta} ($j_{R}-j_{L}=2$)
positions, $\phi \equiv \pm 2\pi /3$. Note that the sign of $\phi $ does not
affect our results. For simplicity we shall assume $|t_{R}/t_{L}|=1$ in the
calculations presented here. Then, the hopping of the leads to the relevant
states of the benzene molecule are $V_{1}^{L}=V_{2}^{L}=V_{1}^{R}=V$, $V_{2}^{R}=Ve^{-i\phi }$, 
with $\phi =0$ for the \emph{para} position, and $\phi =\pm 2\pi /3$ 
for the other two possibilities of connecting the leads.

For $\phi =0$, the state $|B\sigma \rangle =(|1\sigma \rangle -|2\sigma
\rangle )/\sqrt{2}$ decouples from the leads and the transport properties of
the system are the same as those for a single level [$|A\sigma \rangle
=(|1\sigma \rangle +|2\sigma \rangle )/\sqrt{2}$] QD connected to he leads,
with hopping $\sqrt{2}V$, which are well known \cite{nca,izum,costi,proetto,vel}.
Another known limit of the model, which cannot be realized in benzene
molecules, but in rings of a number of sites multiple of four \cite{mole} is 
$\phi =\pi $. In this case, $|A\sigma \rangle $ ($|B\sigma \rangle $) is
coupled only to the left (right) lead and the model is equivalent to an
SU(4) Anderson model \cite{desint,fcm}. However, in contrast to the case of C
nanotubes \cite{lim,ander,lipi,buss} and Si fin-type field effect
transistors \cite{tetta}, the conductance vanishes identically as a
consequence of perfect destructive interference \cite{desint,izum2}.
An important technical difference
is that in our model, the hybridization matrices of the states $|i\sigma
\rangle $ with the left and right leads are not proportional for $\phi \neq 0 $.
If the
degeneracy between the levels can be lifted (for example applying an
external flux \cite{rinc}), the symmetry is reduced to SU(2) and a
finite conductance is restored \cite{desint,izum2,fcm}.
In C nanotubes with disorder induced valley mixing \cite{grove}, the effective model 
becomes very similar to ours, including a finite level splitting
and a phase $\phi$, which depends on the applied magnetic field.

We shall use our previous
results for the case $\phi =\pi $ and a finite level splitting $\delta
=E_{2}-E_{1}$ to help in the analysis of the results presented here for
benzene.

\section{The formalism}
\label{forma}

In this section we describe the extension of the non-crossing approximation
(NCA) applied before for the Anderson model with infinite on-site repulsion out of
equilibrium \cite{nca,nca2}, to our effective Hamiltonian.

\subsection{Representation of the Hamiltonian with slave particles}
\label{repre}

An auxiliary boson $b$, and four auxiliary fermions $f_{i\sigma }$ are
introduced, so that the localized states are represented as

\begin{equation}
|s\rangle =b^{\dag }|0\rangle {\rm , }|i\sigma \rangle =f_{i\sigma }^{\dag
}|0\rangle ,  \label{rep}
\end{equation}%
where $|0\rangle $ is the vacuum. These pseudoparticles should satisfy the
constraint 
\begin{equation}
b^{\dag }b+\sum_{i\sigma }f_{i\sigma }^{\dag }f_{i\sigma }=1.  \label{cons}
\end{equation}%
Introducing it by a Lagrange multiplier, the effective Hamiltonian takes
the form 
\begin{eqnarray}
H^{\prime } &=&(E_{s}+\lambda )b^{\dag }b+\sum_{i\sigma }(E_{i}+\lambda
)f_{i\sigma }^{\dag }f_{i\sigma }+\sum_{\nu k\sigma }\epsilon _{\nu k}c_{\nu
k\sigma }^{\dagger }c_{\nu k\sigma }  \nonumber \\
&&+\sum_{k\nu i\sigma }\left( V_{i}^{\nu }f_{i\sigma }^{\dagger }bc_{k\nu
\sigma }+{\rm H.c.}\right) .  \label{hp}
\end{eqnarray}%
The NCA solves a system of self-consistent equations to obtain the Green
functions of the pseudoparticles (described below), which is equivalent to
sum an infinite series of diagrams (all those without crossings) in the
corresponding perturbation series in the last term of $H^{\prime }$, and
afterwards project on the physical subspace of the constraint. The main
new difficulties compared to the case of one doublet arise as a consequence
of the matrix structure of the pseudofermion Green functions and self
energies, which are absent in the SU(4) case or in this case with simple
SU(2) symmetry breaking perturbations, as a magnetic \cite{tetta} or
"orbital" \cite{desint,izum2,fcm} field.

\subsection{Green functions}
\label{green}

The lesser and greater Keldysh Green functions for the psedoparticles for
stationary non-equilibrium processes are defined as \cite{lif,mahan} 
\begin{eqnarray}
G_{ij,\sigma }^{<}(t-t^{\prime }) &=&+i\langle f_{j\sigma }^{\dag
}(t^{\prime })f_{i\sigma }(t)\rangle ,  \nonumber \\
D^{<}(t-t^{\prime }) &=&-i\langle b^{\dag }(t^{\prime })b(t)\rangle ,  \nonumber
\\
G_{ij,\sigma }^{>}(t-t^{\prime }) &=&-i\langle f_{i\sigma }(t)f_{j\sigma
}^{\dag }(t^{\prime })\rangle ,  \nonumber \\
D^{>}(t-t^{\prime }) &=&-i\langle b(t)b^{\dag }(t^{\prime })\rangle ,
\label{glg}
\end{eqnarray}%
the retarded and advanced fermion Green functions are $G_{ij,\sigma
}^{r}(t)=\theta (t)[G_{ij,\sigma }^{>}(t)+G_{ij,\sigma }^{<}(t)]$, 
$G_{ij,\sigma }^{a}=G_{ij,\sigma }^{r}+G_{ij,\sigma }^{<}-G_{ij,\sigma }^{>}$,
and similarly for the boson Green functions replacing $G_{ij,\sigma }$
by $D$.

These Green functions correspond to the interacting (dressed) propagators
and are determined selfconsistently within the NCA. Evaluating the
corresponding diagrams in second order in the $V_{i}^{\nu }$, and replacing
the bare propagators by the dressed ones, the expressions for the lesser
self-energies take the form 
\begin{eqnarray}
\Pi ^{<}(\omega ) &=&-\sum_{\nu lm\sigma }\Gamma _{lm}^{\nu }\int \frac{%
d\omega ^{\prime }}{2\pi }(1-f_{\nu }(\omega ^{\prime }-\omega
))G_{ml,\sigma }^{<}(\omega ^{\prime }),  
\nonumber \\
\Sigma _{lm,\sigma }^{<}(\omega ) &=&-\sum_{\nu }\Gamma _{lm}^{\nu }\int 
\frac{d\omega ^{\prime }}{2\pi }f_{\nu }(\omega -\omega ^{\prime
})D^{<}(\omega ^{\prime }),  
\label{sigl}
\end{eqnarray}
where

\begin{equation}
\Gamma _{ij}^{\nu }(\omega )=2\pi 
\sum_{k}V_{i}^{\nu }\bar{V}_{j}^{\nu}\delta (\omega -\epsilon _{\nu k})  \label{gam}
\end{equation}
assumed independent of $\omega $. Similarly, the greater self-energies
become 
\begin{eqnarray}
\Sigma _{lm,\sigma }^{>}(\omega ) &=&\sum_{\nu }\Gamma _{lm}^{\nu }\int 
\frac{d\omega ^{\prime }}{2\pi }(1-f_{\nu }(\omega -\omega ^{\prime
}))D^{>}(\omega ^{\prime }),  \nonumber \\
\Pi ^{>}(\omega ) &=&\sum_{\nu lm\sigma }\Gamma _{lm}^{\nu }\int \frac{%
d\omega ^{\prime }}{2\pi }f_{\nu }(\omega ^{\prime }-\omega )G_{ml,\sigma
}^{>}(\omega ^{\prime }).  \label{sigg}
\end{eqnarray}

As in the case of one doublet only \cite{nca}, in the expressions for the
retarded self energies, $\Sigma ^{r}(t)=-\theta (t)(\Sigma ^{<}(t)-\Sigma
^{>}(t))$, $\Sigma ^{<}$ can be neglected in comparison with $\Sigma ^{>}$
due to the effect of the constraint. Then after Fourier transforming one
obtains%
\begin{eqnarray}
\Sigma _{ij,\sigma }^{r}(\omega ) &=&i\int \frac{d\omega ^{\prime }}{2\pi }%
\frac{\Sigma _{ij,\sigma }^{>}(\omega ^{\prime })}{\omega -\omega ^{\prime
}+i\eta },  \nonumber \\
\Pi ^{r}(\omega ) &=&i\int \frac{d\omega ^{\prime }}{2\pi }\frac{\Pi
^{>}(\omega ^{\prime })}{\omega -\omega ^{\prime }+i\eta },  \label{sigr}
\end{eqnarray}%
where $\eta $ is a positive infinitesimal. The advanced self energies $%
\Sigma _{ij,\sigma }^{a}$ and $\Pi ^{a}$ are obtained changing the sign of $%
\eta $ in the expressions above.

For the case of SU(4) symmetry or when this symmetry is broken by simple
symmetry breaking fields \cite{desint,izum2,fcm,lim,ander,lipi,buss,tetta} ($\phi
=\pi $ and symmetric coupling to the leads in our effective model), the
fermion Green functions \ and self energies are diagonal in an appropriate
basis and the self-consistency problem simplifies considerably. In the
general case (including $\phi \equiv \pm 2\pi /3$ for benzene) one has to
solve a matrix Dyson equation \cite{lif,mahan} which includes not only the
Keldysh (or +, - branch index in the Keldysh contour), but also the doublet
index $i=1,2$ in the fermion case. For the retarded fermion Green
functions and self energies, combining $G_{ij,\sigma }^{r}$ in a $2\times 2$
matrix $\mathbf{G}^{r}$ and similarly for self energies and unperturbed
Green functions $g_{ij,\sigma }^{r}=\delta _{ij}(\omega -\tilde{E}_{i})^{-1}$, 
with $\tilde{E}_{i}=E-i+\lambda$ the
Dyson equations take the simple form $\mathbf{G}^{r}=\mathbf{g}^{r}
+\mathbf{g}^{r}\mathbf{\Sigma }^{r}\mathbf{G}^{r}$. Solving the system for the $%
G_{ij,\sigma }^{r}$ we obtain 
\begin{eqnarray}
G_{11,\sigma }^{r} &=&\frac{1}{\mathbb{D}}(\omega -\tilde{E}_{2}-\Sigma _{22,\sigma
}^{r}),  \nonumber \\
G_{12,\sigma }^{r} &=&\frac{1}{\mathbb{D}}(\Sigma _{12,\sigma }^{r}),  \nonumber
\\
G_{21,\sigma }^{r} &=&\frac{1}{\mathbb{D}}(\Sigma _{21,\sigma }^{r}),  \nonumber
\\
G_{22,\sigma }^{r} &=&\frac{1}{\mathbb{D}}(\omega -\tilde{E}_{1}-\Sigma _{11,\sigma
}^{r}),  \label{dysonf}
\end{eqnarray}%
where

\begin{equation*}
\mathbb{D}=(\omega -\tilde{E}_{1}-\Sigma _{11,\sigma }^{r})(\omega -\tilde{E}_{2}-\Sigma
_{22,\sigma }^{r})-\Sigma _{12,\sigma }^{r}\Sigma _{21,\sigma }^{r}.
\end{equation*}

For the boson, which has no subscripts, one has 
\begin{equation}
D^{r}(\omega )=\frac{1}{\omega -\tilde{E}_{s}-\Pi ^{r}}.  \label{dysonb}
\end{equation}

The advanced Green functions can be obtained from the retarded ones by the
replacement $\eta \rightarrow -\eta $.

The remaining equations that relate the lesser and greater pseudoparticle
Green functions with the corresponding self energies are (in compact matrix
form for the fermions) \cite{note}
\begin{eqnarray}
\mathbf{G}^{\lessgtr } &=&\mathbf{G}^{r}\mathbf{\Sigma }^{\lessgtr }\mathbf{G}^{a},  \nonumber \\
D^{\lessgtr } &=&D^{r}\Pi ^{\lessgtr }D^{a}.  \label{dysonne}
\end{eqnarray}

We solve numerically the system of integral equations (\ref{sigl}) to 
(\ref{dysonne}) for the pseudoparticle Green functions and self energies. The details 
are given in appendix \ref{deta}

For the calculation of the current, one needs the Green functions of the
physical fermions $d_{i\sigma }^{\dagger }=|i\sigma \rangle \langle
s|=f_{i\sigma }^{\dagger }b$. These functions, which we identify with the
subscript $\mathbf{d}$, are defined in the same way as the pseudofermion
ones [see Eqs. (\ref{glg})], replacing $f_{i\sigma }$ by $d_{i\sigma }$. In
terms of the auxiliary-particle Green functions, the lesser and greater physical 
Green functions take the form \cite{nca}
\begin{eqnarray}
G_{\mathbf{d}ij,\sigma }^{<}(\omega ) &=&i\int \frac{d\omega ^{\prime }}{%
2\pi Q}G_{ij,\sigma }^{<}(\omega ^{\prime }+\omega )D^{>}(\omega ^{\prime }),
\nonumber \\
G_{\mathbf{d}ij,\sigma }^{>}(\omega ) &=&i\int \frac{d\omega ^{\prime }}{%
2\pi Q}G_{ij,\sigma }^{>}(\omega ^{\prime }+\omega )D^{<}(\omega ^{\prime }),
\label{gd}
\end{eqnarray}%
where $Q=\langle b^{\dag }b+\sum_{i\sigma }f_{i\sigma }^{\dag }f_{i\sigma
}\rangle $ for a given Lagrange multiplier $\lambda $ (see appendix \ref{deta}).

\subsection{Equation for the current}
\label{curre}

Using general formulas for the current through a region with interacting
electrons \cite{meir,jau} and the relation $G^{r}-G^{a}=G^{>}-G^{<}$ between
Green functions, the current of our effective model for a spin degenerate
system (without applied magnetic field) can can be written as 
\begin{eqnarray}
I &=&\pm \frac{ie}{h}\int d\omega {\rm Tr}[(\mathbf{\Gamma ^{L}}%
f_{L}(\omega )-\mathbf{\Gamma ^{R}}f_{R}(\omega ))\mathbf{G}_{\mathbf{d}%
}^{>}(\omega )  \nonumber \\
&+&((1-f_{L}(\omega ))\mathbf{\Gamma ^{L}}-(1-f_{R}(\omega ))\mathbf{\Gamma ^{R}})\mathbf{G}_{%
\mathbf{d}}^{<}(\omega )],  \label{ia}
\end{eqnarray}%
where the + (-) sign correspond to the case in which the doublets have one
more electron (hole) than the singlet, $f_{\nu }(\omega )=[\exp [(\omega -\mu
_{\nu })/kT]+1]^{-1}$ where $\mu _{\nu }$ is the chemical potential of the lead $\nu $,  
and $\mathbf{\Gamma ^{\nu }}$, $\mathbf{%
G}_{\mathbf{d}}^{<}$ and $\mathbf{G}_{\mathbf{d}}^{>}$ are $2\times 2$
matrices with matrix elements given by Eqs (\ref{gam}) and (\ref{gd}). In
particular, taking the unperturbed density of 
conduction states per spin $\rho =1/(2D)$ where $2D$
is the band width, and symmetric coupling to the leads, we have 
\begin{eqnarray}
\mathbf{\Gamma ^{L}} &=&\frac{\pi V^{2}}{D}\left( 
\begin{array}{cc}
1 & 1 \\ 
1 & 1%
\end{array}%
\right) ,  \nonumber \\
\mathbf{\Gamma ^{R}} &=&\frac{\pi V^{2}}{D}\left( 
\begin{array}{cc}
1 & e^{i\phi } \\ 
e^{-i\phi } & 1%
\end{array}%
\right) .  \label{gam2}
\end{eqnarray}%
Note that unless $\phi =0$, $\mathbf{\Gamma ^{L}}$ is not proportional to 
$\mathbf{\Gamma ^{R}}$. As a consequence, the trick used to relate the
conductance at $V_b=0$ with the spectral density of states \cite{meir,jau} cannot
be used. We calculate the conductance $G=dI/dV_b$ by numerical differentiation
of the current even for $V_b \rightarrow 0$. The traces appearing in Eq. (\ref{ia}) have the form

\begin{eqnarray}
{\rm Tr}(\mathbf{\Gamma ^{R}G}_{\mathbf{d}}^{\lessgtr }) &=&\frac{\pi V^{2}%
}{D}[G_{11}^{\lessgtr }+G_{22}^{\lessgtr }+\cos (\phi )(G_{21}^{\lessgtr
}+G_{12}^{\lessgtr })  \nonumber \\
&&+i~{\rm sin}(\phi )(G_{21}^{\lessgtr }-G_{12}^{\lessgtr }),  
\label{tr1}
\end{eqnarray}%
and similarly for Tr$(\mathbf{\Gamma ^{L}G}_{\mathbf{d}}^{\lessgtr })$
replacing $\phi $ by 0.

Note that from the definition of the Green functions [see Eq. (\ref{glg})],
one realizes that the complex conjugates of the lesser and greater 
Green functions satisfy 
$\bar{G}_{ij}^{\lessgtr }(t)=$ -$G_{ji}^{\lessgtr }(-t)$ and after Fourier
transform $\bar{G}_{ij}^{\lessgtr }(\omega )=$ -$G_{ji}^{\lessgtr }(\omega )$. 
Then, $G_{ii}^{\lessgtr }(\omega )$ and $G_{21}^{\lessgtr }(\omega
)+G_{12}^{\lessgtr }(\omega )$ are pure imaginary, while $G_{21}^{\lessgtr
}(\omega )-G_{12}^{\lessgtr }(\omega )$ is real.

In appendix A we show that the current from the left lead to the molecule
equals that from the molecule to the right lead, so that the current is
conserved within the NCA. Other approximations, like perturbation theory in
the Coulomb repulsion out of the electron-hole symmetric point do not
conserve the current \cite{scali}, unless some tricks are used \cite{levy,none,mon}.

\section{Numerical results}
\label{res}

For the numerical calculations, we assume a constant density of states per
spin of the leads $\rho $ between $-D$ and $D$. We take the unit of energy
as the total level width of both doublets:  $\Gamma =\Gamma _{ii}^{L}+\Gamma
_{ii}^{R} =4\pi \rho V^{2}$, $i=1,2$. Also $\Gamma =2\Delta $, where 
$\Delta $, called the resonance level width, is half the width at half
maximum of the spectral density of states of particles for each level in the
non-interacting case. We restrict our present study to gate voltages such
that $E_{s}-E_{i}\gg \Delta $, for which correlations play a more important
role, the Kondo effect develops and the conductance is higher. Without loss
of generality, we take $\epsilon _{F}=E_{s}=0$, where $\epsilon _{F}$ is the
Fermi level of the leads without applied bias voltage. 
We define $E_d=E_1$, $\delta=E_2-E_1$.
For a bias voltage  $V_b$, 
the chemical potentials are $\mu _{L}=eV_b/2$, $\mu _{R}=-eV_b/2$.

In the numerical results an important low-energy scale is the Kondo temperature 
$T_K$. It is known that for impurity Anderson models like ours for $\delta=0$ in the Kondo regime,
the spectral density shows one charge-transfer peak at energy $E_d$ below the Fermi energy of
total width of the order of $N \Gamma$ for SU(N) models \cite{fcm,lim,logan}. For finite on-site Coulomb
repulsion $U$ there is another charge-transfer peak at energy $E_d+U$, but this is
shifted to infinite energies in our case. In addition, there is a narrow peak near the Fermi
energy of width of the order of $T_K$. In this work we define $T_K$ as the half width
at half maximum of this peak. For finite $\delta$ additional peaks appear as shown in Section \ref{spec}.

\subsection{Conductance out of equilibrium}
\label{condu}

\begin{figure}[tbp]
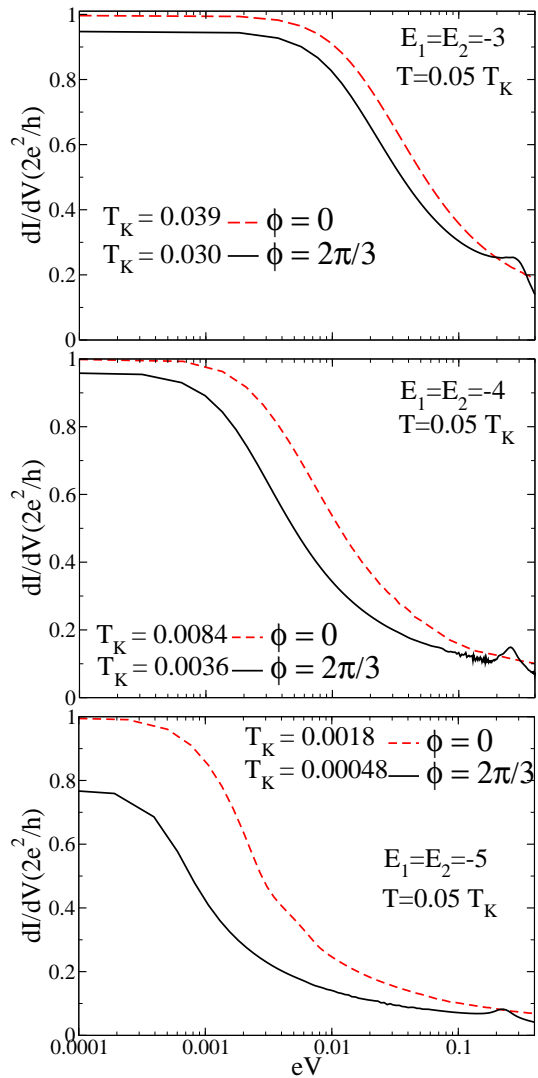

\includegraphics[width=7.0cm]{fig2a.eps} \\
\includegraphics[width=7.0cm]{fig2b.eps} \\ 
\includegraphics[width=7.0cm]{fig2c.eps} \\ 
\caption{(Color online) Differential conductance at low temperature through a benzene molecule as a function of bias voltage 
for several values of the energy of the doublets, with the leads connected in the \emph{para} (red dashed line)
and \emph{ortho} or \emph{meta} (full black line) positions.}
\label{didv}
\end{figure}

 In Fig. \ref{didv} we show the differential conductance $G=dI/dV�_b$ as a function of bias voltage $V_b$, for
the leads connected at 180 degrees (in the \emph{para} position), for which $\phi =0$, and in the 
other two positions ($\phi = \pm 2\pi /3$), for several values of the energy of the doublets $E_1=E_2=E_d$.
The charge-transfer energy $\epsilon_F-E_d$  is the energy necessary to take a particle (electron or hole) from the Fermi energy, 
and add it to the localized singlet, forming a doublet, in absence of the hopping to the leads. It is tuned modifying 
the gate voltage $V_g$: $E_d$ decreases with increasing $V_g$. The temperature in the curve was fixed at $T=0.05 T_K$.
Since we have assumed symmetric coupling to both leads, $G(-V_b)=G(V_b)$ and only positive $V_b$ are du¡isplayed in the figure. 

 For the \emph{para} position, as explained in Section \ref{didv}, the problem is equivalent to the transport through a single level
quantum dot, and $G$ shows a single peak centered near zero voltage and width of the order of $2 k T_K/e$ \cite{fcm2}. We remind the 
reader that half the width of $G(V_b)$ for $T=0$ times the electric charge $e$, half the width of the spectral density and the 
temperature (times the Boltzman constant) $T$ at which $G(T)$ for $V_b=0$ falls to half of the maximum value, are all of the 
same order \cite{fcm2}. This is natural from the expected universal behavior of the (single level) Anderson model in the Kondo regime, 
in which only one energy scale $T_K$ survives. This Kondo temperature varies exponentially with the charge-transfer energy $E_d$, 
and as a consequence, the width of the central peak observed in the figure also decreases exponentially with increasing $|E_d|$, 
as observed in Fig. \ref{didv}

 When the molecule is attached to the leads at 60 or 120 degrees, three main changes occur in $G(V_b)$ in comparison to the previous case:
i) The maximum of $G(0)$ is lower, ii) The width of the central peak is narrower and scales in a different way with $E_d$, and iii) 
two new peaks appear at finite $V_b$. 
 i) is due to the effect of partial destructive interference between the current transmitted by the
two doublets. In the \emph{para} position, for which all hybridization to the leads have the same sign ($\phi =0$), 
$G(0)$ equals the quantum of conductance $2 e^2/h$ for symmetric coupling to the leads 
(as we have assumed in the present calculations). In the opposite case (hypothetical for 
benzene) of $\phi = \pi$, there is perfect destructive interference and $G$ vanishes \cite{desint,izum2}. Transport through the 
benzene molecule connected in the \emph{ortho} or \emph{meta} positions, correspond to an intermediate situation 
($\phi = \pm 2\pi /3$), and therefore, an intermediate value of $G(0)$ is expected.  For asymmetric coupling to the leads, 
as it is usually the case in transport through molecules in devices built by electromigration \cite{roch,serge}, 
$G(0)$ decreases strongly and it is not possible to distinguish between different positions of the leads connected to the 
benzene molecule from the maximum value of $G$. Note that this value decreases slightly with decreasing $T_K$. The feature,
which is probably the most useful to distinguish experimentally among the different positions of the conducting leads 
is the presence of the side peaks
in $G(V_b)$, which seem to depend weakly on the charge transfer energy $\epsilon_F-E_d$, 
in contrast to the width of the central peak.

\begin{figure}[tbp]
\includegraphics[width=7.0cm ]{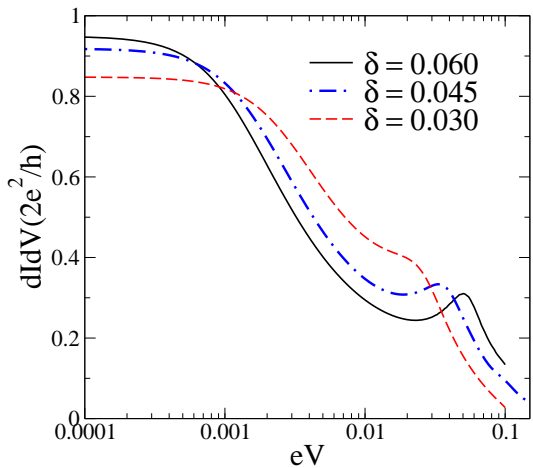}
\caption{(Color online) Differential conductance for $\phi=\pi$ and different values of the level splitting.
Other parameters are $E_d=-4$, $T=0.05 T_K$.}
\label{didvpi}
\end{figure}

In order to help in the interpretation of the results, we calculate $G(V_b)$ for $\phi=\pi$, 
but allowing for a finite splitting $\delta=E_2-E_1$ between both doublets. $\delta$ 
acts as a symmetry breaking field on the SU(4) Kondo effect \cite{desint,fcm}.
The result is shown in Fig. \ref{didvpi}. Clearly, $G(V_b)$ is qualitatively similar to the corresponding 
result for benzene with leads connected in the \emph{ortho} or \emph{meta} positions. 
The conductance, which vanishes in the SU(4) limit $\delta=0$ is restored by a finite $\delta$
and two peaks at $eV_b= \pm \delta$ appear. 
 This similarity suggest to interpret the
results for benzene starting from those for $\phi=\pi$, $\delta=0$ and thinking the difference 
between the coupling $V^R_2$ for $\phi=\pm 2 \pi/3$ and $\phi=\pi$ as a perturbation. This
perturbation, in second order, introduces among other effects, an effective mixing between 
the levels, proportional to $V^2/|E_d|$, which leads to a splitting of the doublets. 
In fact, the position of the satellite peaks in Fig. \ref{didv} (ranging from 0.3 for $E_d=-3$ 
to 0.22 for $E_d=-5$) is roughly consistent with a $1/|E_d|$ dependence. 
 
Note that in the case $\phi=0$ in which this perturbation is expected to be the largest, only
the symmetric combinations of left and right lead states hybridize with the impurity, 
while the antisymmetric ones remain decoupled. The spectral density of the latter 
is a delta function $\delta(\omega-E_d)$, without weight near the Fermi energy.
This makes clear that actually the effective hopping of the two resulting split doublets is different
and the proposed interpretation is rather qualitative.

\subsection{Spectral densities}
\label{spec}

\begin{figure}[tbp]
\includegraphics[width=7.0cm ]{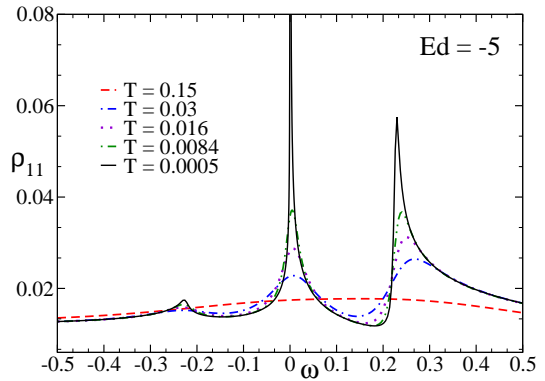}
\caption{(Color online) Equilibrium spectral density for the benzene doublets as a function of frequency in the 
\emph{ortho} and 
\emph{meta} positions for different temperatures.}
\label{rho}
\end{figure}

\begin{figure}[tbp]
\includegraphics[width=7.0cm ]{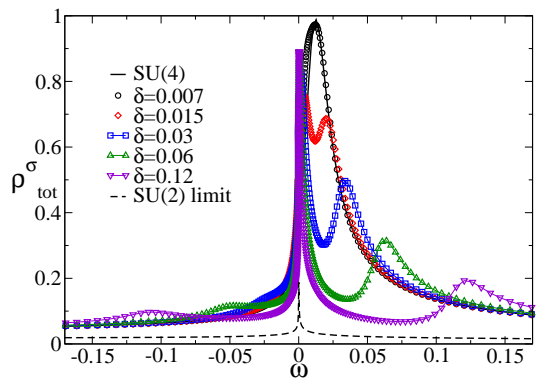}
\caption{(Color online) Total spectral density at equilibrium as a function of frequency for the effective model 
with $\phi=\pi$, $T=0.05 T_K$ and different values of the level splitting.}
\label{rhopi}
\end{figure}

As a further test of the  
interpretation outlined above, we compare the spectral density of states of both models. 

The spectral density
\begin{equation}
\rho_{i \sigma} (\omega )=(G_{{\bf d}ii \sigma}(\omega + i\eta) - G_{{\bf d}ii \sigma}(\omega - i\eta))/(2 \pi i)
\label{rhos}
\end{equation}
of both doublets for benzene connected to the leads in \emph{ortho} or \emph{meta} positions
is represented in Fig. \ref{rho} for different temperatures and energies near the Fermi 
energy (the charge-transfer peak \cite{fcm} is not shown). As $T$ decreases below $T_K$,  not only a
peak develops near the Fermi energy, but also two side peaks 
(the most prominent for positive frequencies) are clearly present. In Fig. \ref{rhopi} we show the total 
spectral density for $\phi=\pi$, 
very low temperatures and different values of $\delta$. In this 
case, the peak near to the Fermi energy corresponds to the lowest doublet and the peak at
energy near $\delta$ corresponds to the highest doublet \cite{fcm}. In contrast, each peak
in Fig. \ref{rho} is expected to come from some linear combination of both doublets which is not easy 
to identify. 
Another difference apparent in the figure is that the side peak at positive frequencies for benzene
is sharper and more asymmetric than the corresponding one for $\phi=\pi$ and finite $\delta$.
This might be due in part to a smaller effective hybridization of the excited doublet in the case of benzene, 
leading to a narrower peak,
as in the limit $\phi \rightarrow 0$ discussed above.

In spite of these differences, the spectral density for $\phi=\pi$ and finite $\delta$ shows the development of a
satellite peak departing from the central one as $\delta$ increases. The spectral density for benzene
shows the same qualitative features (a peak near the Fermi energy and another one at positive frequencies)
which can be interpreted, in a qualitative first approximation, as coming from an effective $\delta$.
This peak at finite energy in turn gives
rise to the side peaks in $G(V_b)$ (although the spectral densities are modified with the
applied bias voltage and the peaks are blurred).

\subsection{Temperature dependence of the conductance}
\label{temper}

\begin{figure}[tbp]
\includegraphics[width=7.0cm ]{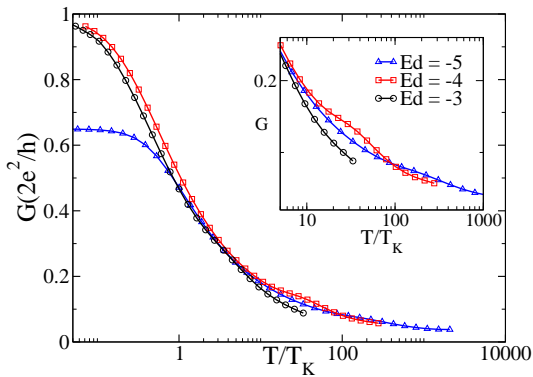}
\caption{(Color online) Conductance as a function of temperature for leads connected in 
the \emph{ortho} or \emph{meta} position and several values of 
the energy of the doublets.}
\label{temp}
\end{figure}

In Fig. \ref{temp}, we show the evolution with temperature of the
conductance $G(T)$ for $V_b=0$, several values of $E_d$, and conducting leads
in the \emph{ortho} or \emph{meta} position.  $G(T)$ was obtained differentiating the current,
as explained in Section \ref{curre}.
In order to be able to display
all curves in the same scale, a logarithmic scale of relative temperatures $T/T_K$ is used,
where the characteristic energy scale $T_K$ is given by the half width at half maximum 
of the equilibrium spectral density $\rho_{i \sigma} (\omega )$. The resulting value
of $T_K$ is $3 \times 10^{-2}$, $3.6 \times 10^{-3}$ and $4.8 \times 10^{-4}$ for $E_d=$ -3, -4 and -5
respectively. The peak at finite energy in the spectral density at low temperatures 
(which lies between 0.3 for $E_d=-3$ to 0.22 for $E_d=-5$) results in a structure 
in $G(T)$ at $T/T_K$ near 10, 70 and 400 for $E_d=$ -3, -4 and -5
respectively. While for $E_d=-3$ this structure is hidden in the main peak, a kind
of plateau is evident for the other values of $E_d$ at the corresponding positions, as shown in the inset of Fig. \ref{temp}.

This structure is reminiscent of the plateau observed by transport experiments in the ``triplet'' side
of the quantum phase transition in C$_{60}$ \cite{roch,serge}, and explained by 
NCA \cite{st1,st2,serge} and NRG calculations with improved resolution \cite{serge},
In our case, to represent accurately the high energy features, it was necessary to use 
a special mesh in the frequency axis with more points near the high energy peak in the 
pseudofermion spectral densities (see appendix \ref{deta}).  

In previous work \cite{desint,fcm}, we have found that for $\phi=\pi$ and any level splitting
$\delta=E_2-E_1$, $T_K$ (except for a constant of order unity) is very 
accurately given by the expression

\begin{equation}
T_{K}=\left\{ (D+\delta )D\exp \left[ \pi E_{1}/(2\Delta )\right]
+\delta ^{2}/4\right\} ^{1/2}-\delta /2,  
\label{tk}
\end{equation}
obtained from a simple variational wave function that generalizes the proposal of Varma and Yafet \cite{varma}
for the simplest impurity Anderson model. This equation interpolates between the SU(4) limit $\delta=0$,
for which $T_K$ has the largest value $T_K^4=D\exp \left[ \pi E_{1}/(4\Delta )\right]$ and the  
one-level SU(2) limit $\delta \rightarrow +\infty$. 
 The effect of $\delta$ on $T_K$ is small when $\delta < T_K^4$, 
while for larger $\delta$, $T_K$ decreases strongly. Note that $T_K^4$ coincides with the Kondo temperature
of the equivalent one-level effective Anderson for benzene connected in the \emph{para} position (because the effective 
one-level resonant level width contains a factor 2, see Section \ref{model}). Therefore, we can interpret
the fact that the effective energy scale in the \emph{ortho} and 
\emph{meta} positions is smaller than that in the \emph{para} positions (compare the widths of the central peaks
in Fig. \ref{didv}) as an effect of the effective level splitting caused by a phase $\phi$ different from $\pi$.
 
In other words, it seems that the position of the side peaks in $G(V_b)$ (or spectral  densities) and the plateau
in $G(T)$ indicate an effective $\delta$, which decreases the characteristic energy scale $T_K$. While we do not expect 
Eq. (\ref{tk}) to describe accurately $T_K$ using this effective $\delta$, the fact that the difference between
the energy scales for the \emph{para} and the other positions is more noticeable for $\delta$ larger 
than $T_K^4$ ($T_K$ for the   
\emph{para} case), suggests that this equation might be useful for a qualitative understanding, and supports 
the interpretation of our results.

\section{Summary and discussion}
\label{sum}

We have constructed the effective Hamiltonian for transport through a symmetric ring 
(point group including $C_{nv}$ as a subgroup) with one orbital per site, including a singlet and two degenerate doublets 
of a neighboring configuration for the isolated ring. 
This includes for example the ground state of the isolated ring for one electron per site, and
another with one added electron or hole. 
The resulting effective generalized Anderson Hamiltonian
describes however more general situations. An extension to the case in which a magnetic flux is added 
(reducing the symmetry to $C_{n}$ and breaking the doublet degeneracy) is trivial. 
Partial destructive interference
occurs in general when two levels with $N$ particles are near to another one
with $N\pm 1$ particles \cite{rinc}. Therefore, the situation of one singlet
and two doublets appears frequently. From the wave vectors of these states,
it is easy to calculate the phase $\phi$ of the effective model following
the lines of Section \ref{model}. Therefore, we expect that our formalism can be used in a variety
of physically relevant systems.

We have used the resulting effective Hamiltonian to calculate the non-equilibrium transport
through a benzene molecule with conducting leads connected in different positions, using
an appropriate generalization of the NCA to include the Kondo effect as well as effects of partial 
destructive interference between the transport channels through both doublets. While 
for leads connected in the \emph{para} position, the conductance $G$ as a function of voltage and 
temperature is similar to that well known for the case of only one doublet, in the other positions 
the peak in $G(V_b)$ near zero bias voltage $V_b$ is narrower and of smaller intensity as an indication of the relative
position of the conducting leads.
In addition, $G(V_b)$ displays two additional side peaks at finite $\pm V_b$, and a characteristic
plateau is present in the conductance as a function of temperature $G(T)$. These finite energy features are probably easier to 
detect experimentally as an indication of the relative position of the conducting leads.

These results for the leads connected in the \emph{ortho} and \emph{meta} positions,
which are due to partial destructive interference, can be interpreted starting from those with
total destructive interference (corresponding to a phase $\phi=\pi$)
for which the effective model has SU(4) symmetry,
and adding a symmetry breaking splitting $\delta$ between effective doublets caused by 
the remaining term treated as a perturbation. 

For our results to be valid, the coupling to the leads should be sufficiently small,
so that excited states, not included in the effective Hamiltonian, do not play an important
role. The effect of these states can be estimated for each particular case. 
See for example Ref. \cite{lobos}. The magnitude of the coupling can be controlled 
for example in break junctions \cite{park,parks,reed}.

Our effective model, including a finite level splitting $\delta$ from the 
beginning, and a general phase $\phi$ describes carbon nanotube QDs 
with disorder induced valley mixing \cite{grove}. Therefore, our formalism 
supplemented by realistic calculations of $\delta$ and $\phi$ might be 
used to calculate the conductance of these systems.

In this work, we have neglected the effect of phonons, which can modify the
effective parameters of the model \cite{har1,pablo}, and also affect the
interference phenomena \cite{har2}. We have also neglected electrostatic
interactions between the leads and the molecules. In the weak coupling regime,
it has been shown that interaction with image charges for the doped molecules leads 
to a symmetry breaking and a splitting of the degenerate levels that also affects the 
interference phenomena \cite{kaa}. As we have shown, in the specific case of
benzene, the coupling to the leads already originates a splitting of this kind. Therefore, we
expect that these electrostatic effects are more relevant in the case of 
perfect destructive interference.

\section*{Acknowledgments}

We thank CONICET from Argentina for financial support. This work was
partially supported by PIP 11220080101821 of CONICET and PICT R1776 of the ANPCyT, Argentina.

\appendix

\section{Current conservation}
\label{cc}

Here we present a brief demonstration that within NCA, the current
established between the left metallic contact and the central region (ring) $%
I_{L}$, is the same as the current flowing from this region to the right
metallic contact $I_{R}$. For simplicity we assume that there is no applied
magnetic field. The demonstration can be easily extended to the case of
inequivalent spins. Adding both spins the expressions of the currents
are \cite{meir} 
\begin{eqnarray}
I_{L} &=&\pm \frac{2ie}{h}\int d\omega \ {\rm Tr} \{ \mathbf{\Gamma ^{L}}%
[f_{L}(\omega )\mathbf{G}_{\mathbf{d}}^{>}(\omega ) \nonumber \\ 
&+&(1-f_{L}(\omega ))
\mathbf{G}_{\mathbf{d}}^{<}(\omega )] \},  \label{il} \\
I_{R} &=&\pm \frac{-2ie}{h}\int d\omega \ {\rm Tr} \{\mathbf{\Gamma ^{R}}%
[f_{R}(\omega )\mathbf{G}_{\mathbf{d}}^{>}(\omega ) \nonumber \\
&+&(1-f_{R}(\omega ))%
\mathbf{G}_{\mathbf{d}}^{<}(\omega )] \},  \label{ir}
\end{eqnarray}
where 
the symbol $\pm $ and the meaning of the matrices is explained
in Section \ref{forma} [actually Eq. (\ref{ia}) is the average of Eqs. (\ref{il}) and
(\ref{ir})].

We use Eqs. (\ref{gd}) to replace the physical Green functions by their
expressions in terms of the Green functions for the auxiliary particles.
Denoting $\mathbf{G}$ the $2\times 2$ matrix with the pseudo fermion Green
functions for a given spin,. one obtains

\begin{eqnarray}
I_{L} &=&\mp \frac{2e}{h}\int d\omega \int \frac{d\omega ^{\prime }}{2\pi Q}%
f_{L}(\omega )D^{<}(\omega ^{\prime })\ {\rm Tr}\left[ \mathbf{\Gamma ^{L}G}%
^{>}(\omega ^{\prime }+\omega )\right]   \nonumber \\
&+&(1-f_{L}(\omega ))D^{>}(\omega ^{\prime })\ {\rm Tr}\left[ \mathbf{%
\Gamma ^{L}G}^{<}(\omega ^{\prime }+\omega )\right]   \label{il2} \\
I_{R} &=&\pm \frac{2e}{h}\int d\omega \int \frac{d\omega ^{\prime }}{2\pi Q}%
f_{R}(\omega )D^{<}(\omega ^{\prime })\ {\rm Tr}\left[ \mathbf{\Gamma ^{R}G}%
^{>}(\omega ^{\prime }+\omega )\right]   \nonumber \\
&+&(1-f_{R}(\omega ))D^{>}(\omega ^{\prime })\ {\rm Tr}\left[ \mathbf{%
\Gamma ^{R}G}^{<}(\omega ^{\prime }+\omega )\right] .  \label{ir2}
\end{eqnarray}%
Using Eqs. (\ref{sigg}) for the  boson self-energies, and some algebra, the
difference between both expressions becomes   
\begin{equation}
I_{L}-I_{R}=\frac{-e}{h}\int \frac{d\omega ^{\prime }}{Q}\left[ D^{<}(\omega
^{\prime })\Pi ^{>}(\omega ^{\prime })-D^{>}(\omega ^{\prime })\Pi
^{<}(\omega ^{\prime })\right] ,  \label{dif}
\end{equation}%
which is easily seen to vanish using $D^{\lessgtr }=D^{r}\Pi ^{\lessgtr
}D^{a}$ [see Eqs. (\ref{dysonne})].

\section{Numerical procedure}
\label{deta}

In this appendix we give some details about the numerical solution of the
system of integral equations. 

We first solve the system that determines the retarded and greater Green
functions and self energies for the auxiliary particles, Eqs. (\ref{sigg})
to (\ref{dysonne}). After the system converges, we solve self-consistenlty
Eqs. (\ref{sigl}) and (\ref{dysonne}) for the lesser quantities. 

Due to the gauge symmetry of the Hamiltonian in the representation of the
auxiliary particles, $f_{i\sigma }\rightarrow \exp (i\lambda
_{0}t)f_{i\sigma }$, $b\rightarrow f_{i\sigma }\exp (i\lambda _{0}t)b$, the
Lagrange multiplier $\lambda $ in Eq. (\ref{gd}) can be shifted by an
arbitrary  number $\lambda _{0}$ \cite{nca2}. This is equivalent to a shift
in the energies of the slave particles or of the frequency as $\omega
\rightarrow \omega +\lambda _{0}$ . Therefore, we can either set $Q=\langle
b^{\dag }b+\sum_{i\sigma }f_{i\sigma }^{\dag }f_{i\sigma }\rangle =1$ and
calculate $\lambda $ at each iteration or fix $\lambda $, calculate $Q$ and
divide by $Q$ in the evaluation of physical properties \cite{nca,nca2}. 

In the present problem, the spectral functions of the pseudo-fermions have
two different peaks in contrast to the single one present in the ordinary
Anderson model. We found that fixing $\lambda $ as the solution of the
equation Re$D(\omega ,\mathbb{\lambda )}=0$ at each iteration, the position
of the lowest peak of the auxiliary fermion spectral densities coincide with
the zero of energy (the Fermi level) With this chice of $\lambda $, in order
to make the numerical evaluations efficiently, we use a non equidistant
frequency mesh $\omega _{i}$ with two different regions logarithmically
spaced. One of them is aimed at the zero of energy $\omega =0$ and the other
one is centered near to the position of the second peak of the fermion
spectral functions, $\omega =\delta $. Finally, to generate the non
equidistant mesh we follow reference \cite{nca2}.


\begin{thebibliography}{99}
\bibitem{note} In the general case, Eqs. (14) contain an additional term of 
the form $|G^r/g^r|^2 g^{\lessgtr }$ (see for example Eq. (7) of Ref. \cite{none}), 
but they vanish in this case because 
$(g_{ij,\sigma }^{r})^{-1}=\delta _{ij}(\omega -\tilde{E}_{i})$
and $g^{\lessgtr } \sim \delta _{ij}\delta(\omega- \tilde{E}_i)$.
\end{thebibliography}
\end{document}